\documentclass[twoside,twocolumn,letterpaper,9pt]{article}
\usepackage{amsmath,bm,amsthm,amsfonts,amssymb,amscd,esint}
\usepackage[left=1.5cm, right=1.5cm, top=1.785cm, bottom=2.0cm]{geometry}
\usepackage{graphicx} 
\usepackage{xcolor}
\usepackage[super,sort&compress,comma,nonamebreak]{natbib} 
\usepackage{balance}
\usepackage{authblk}

\usepackage{float}
\usepackage{framed}
\usepackage{xcolor}
\colorlet{shadecolor}{orange!15}

\def\dd{{\rm d}} 
\def\br{{\bf r}}
\def\er{\epsilon_{\rm r}}
\def\ein{\epsilon_{\rm in}}

\def\eout{\epsilon_{\rm out}}

\def\hatbn{\hat{\bf n}}

\def\tldL{\hbox{$\bf \tilde{\phantom{L}}$\kern-.55em\hbox{$\bf L$}}}
\def\rhof{\rho\lower3pt\hbox{${}_{\rm f}$}}

\title{Exact polarization energy for clusters of contacting dielectrics}
\author[1]{Huada Lian}
\author[2,*]{Jian Qin}
\affil[1]{Department of Materials Science \& Engineering, Stanford University}
\affil[2]{Department of Chemical Engineering, Stanford University}
\affil[*]{Corresponding author: jianq@stanford.edu}
\date{\today}

\begin{document}

\twocolumn[
\begin{@twocolumnfalse}
\maketitle
\begin{abstract}
The induced surface charges appear to
diverge when dielectric particles form close contacts.
Resolving this singularity numerically
is prohibitively expensive because high spatial resolution is needed.
We show that the strength of this singularity is
logarithmic in both inter-particle separation and
dielectric permittivity.
A regularization scheme is proposed
to isolate this singularity,
and to calculate the exact cohesive energy
for clusters of contacting dielectric particles.
The results indicate that polarization energy
stabilizes clusters of open configurations
when permittivity is high,
in agreement with the behavior of conducting particles,
but stabilizes the compact configurations when permittivity is low. 
\end{abstract}
\end{@twocolumnfalse}
]

\noindent
For particle aggregates or clusters
stabilized by electrostatic
interactions,\cite{kolehmainen2018effects,
lee2015direct,shevchenko2006structural}
the cohesive energy 
depends on the dielectric permittivities
of both the particles and the medium.
Permittivity quantifies the density of
dipoles induced by externally applied electric fields,
which is proportional to polarization
in the linear regime.
When the permittivity contrast between the particles and the medium is high,
as is often the case,
the polarizations from the two sides of the interface
do not fully compensate each other,
resulting in the accumulation of induced surface charges.

Resolving the surface charges is needed
to evaluating the electrostatic interactions
among particles in close proximity,
and is challenging because polarization
is intrinsically a many-body effect,
depending on the positions of all particles.
For instance, careful measurements
in colloidal suspensions have shown
that the inter-particle force is non-additive,
which is at least partially due to the polarization effect.\cite{merrill2009many}
In another set of experiments on metallic nanoparticles,
it has been found that the aggregation of multiple particles
surrounding a charged particle can be stabilized
by the polarization effect alone.\cite{shevchenko2006structural,qin2016singular}
More dramatic demonstration of such polarization effects 
is found in the so-called like-charge attraction
caused by the strong polarization effect,
for particles of large size or permittivity ratios.\cite{russell1909coefficients, lekner2012electrostatics}

Computational methods,
such as the boundary element method,\cite{barros2014efficient}
the spectral methods~\cite{lindgren2018integral,lian2018polarization} 
and the image method,\cite{qin2016image} 
have been developed to account for this polarization effect.
In practice, these methods all need
to evaluate the induced surface charge densities in one form or another,
and have been successfully applied to study a wide range
of problems involving the aggregation of polarizable particles.\cite{shen2017surface,barros2014dielectric,sherman2018field}
However, when particles are in close proximity,
the surface charge densities apparently diverge
because the electric field in the narrow gap region is strong
even for a small difference in the electrostatic potentials of particles.
This is analogous to the divergent lubrication force
between approaching solid particles in the Stokesian regime.\cite{leal2007advanced}
Resolving these apparently diverging charge densities
requires a high spatial resolution
that becomes computationally prohibitive
for nearly touching particles.\cite{barros2014efficient}
For conductors like metalic particles,
the functional form of the divergent charge density has been identified
and used to isolate the singularity obscuring numerical calculations,
and to obtain the exact energy and force for
contacting particles.\cite{qin2016singular}
For dielectric particles, the question remains unsolved.

\begin{figure}[b!]
    \centering
    \includegraphics[width=0.49\textwidth]{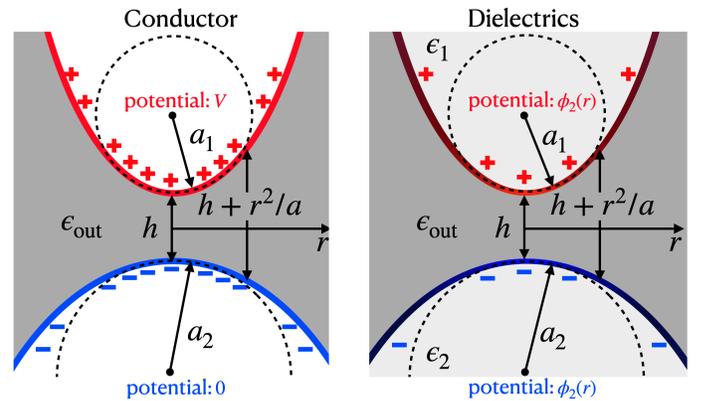}
    \caption{
    Surface charges between two particles 
    separated by a distance~$h$.
    The separation of two surfaces at a radial distance $r$ 
    from the contact points can be approximated by $h+r^2/a$,
    where $a\equiv2/(a_1^{-1}+a_2^{-1})$. 
    The permittivity of the medium is $\eout$. 
    Surface of conducting particles are equipotential 
    with potential set to $V$ and $0$.
    Surfaces of dielectric particles, 
    with permittivities $\epsilon_1$ and $\epsilon_2$, 
    have surface potential distributions $\phi_1(r)$ and $\phi_2(r)$, respectively. 
    }
    \label{fig:gap}
\end{figure}

One way to reveal these singular surface charges
is to consider two conducting particles
separated by a small gap,
as sketched in Fig.\,\ref{fig:gap}. 
The points on the two surfaces
separated by the minimum distance, or gap distance~$h$,
are referred to as {contact points}.
The surface charge density 
is proportional to the strength 
of the normal component of electric field 
on surfaces, according to Gauss's law. 
When $h$ is small, 
the electric field lines in the gap region
are nearly parallel to
the line connecting the two contact points.
The strength of the electrical field is, in turn,
proportional to the difference in the surface electrical potentials.
This relation between the surface potential 
and the electrical field
is analogous to
that between the longitudinal velocities of converging particles 
and the transverse velocity of squzzed fluid velocity
in the lubrication theory.\cite{leal2007advanced}

To calculate the surface charge density,
both the surface potentials and the vertical separations are needed.
Since conductors are equipotential,
the surface potentials
can be set to~$V$ and~$0$ respectively.
The vertical separation depends on the radial distance~$r$,
and can be expressed within the Derjaguin approximation
as $h+r^2/a$,
where $a$ is the harmonic average of 
radii of curvature $a_1$ and $a_2$ 
at the two contact points,
i.e., $a^{-1}=(a_1^{-1}+a_2^{-1})/2$. 
Here, the spherical apexes are assumed for simplicity,
but more general curvatures
can be treated similarly.
Consequently, the field strength 
at radius $r$ is $E(r) = V/(h+r^2/a)$, 
and the surface charge density 
$\sigma(r)=\frac{1}{4\pi}\eout V/(h+r^2/a)$
on the top surface,  
where~$\eout$ is the medium permittivity.
Integrating $\sigma(r)$ for $r$ from $0$ to $R_0$, 
where $R_0$ is a cutoff of order $a$,
results in the singular part of the surface charge 
\begin{equation}
    \begin{aligned}
    Q_{\rm s} &= \frac{V \eout }{4\pi}
    \int_0^{R_0}\! \dd r 
    \frac{2\pi r }{h+r^2/a} 
    \simeq \frac{V\eout a}{4}  \ln\left(\frac ah\right) .
    \end{aligned}
    \label{eq:QsingularC}
\end{equation}
The unity in the logarithmic term is dropped
because $R_0\simeq a\gg h$.
Further, the difference between~$R_0$ and~$a$ is neglected
because it only leads to a constant shift.
Equation~(\ref{eq:QsingularC}) shows how surface charges
of the upper surface
become singular as the gap distance decreases.
That of the lower surface
is also singular, but of opposite sign.
In the limit~$h \to 0$,
the energy does not blow up 
because the potential difference $V$ vanishes
once particles form contact.

The ratio of the singular charge $Q_{\rm s}$ and the potential difference $V$
gives the singular part of the capacitance~$c_{\rm s} = \eout a \ln(a/h)/4$,
which has been found previous for spherical dimers.\cite{russell1909coefficients,lekner2012electrostatics}
The above analysis 
shows that this singular capacitance
is local.
Thus, the same singularity
applies to non-spherical particles,
and for aggregates of multiple particles.
This fact has been employed to find the exact energy
for clusters of contacting, conducting particles.\cite{qin2016singular}
In this work, the full capacitance array
is first numerically calculated
for an ensemble of conducting particles at finite
but small separations.
The singular contribution is then subtracted,
leaving a regular part that can be extrapolated to~$h = 0$.
Finally, when the singular and regular parts are pieced together,
the variation of energy with separation is obtained.

For the dielectric case,
a straightforward generalization of the above treatment fails,
because the particles are not equipotential.
The potential difference~$\Delta V(r)$ needed to
evaluate the charge density in eq.\,(\ref{eq:QsingularC})
can not be fully specified by the potential difference
between the contact points~$\Delta V(0)$.
The variation of~$\Delta V(r)$ with~$r$ in the gap region
is expected to be quadratic, but the curvature is unknown {\it a priori}.
In an early, and analogous work on thermal conduction of composite
materials, Batchelor et al.\cite{batchelor1977thermal} noticed
that the potential distribution, which determines the surface charges,
is itself dominated by the contribution from the surface charges nearby,
so that a self-consistent treatment is needed.

To illustrate this point, we consider
the dielectric case
in Fig.\,\ref{fig:gap}.
Let the surface potentials of the two particles be
$\phi_1(r)$ and $\phi_2(r)$,
and the surface charges be~$\sigma_1(r)$ and~$\sigma_2(r)$.
The surface potential $\phi_i(r)$,
with $i = 1, 2$,
can be calculated by integrating
the coulombic potential of the respective surface charges.
Near the contact point, we have
\begin{equation}
   \phi_i(r) =
   \tilde{V_i} 
   +\frac{1}{2\pi\epsilon_i}
   \int_0^\infty\!\!\dd r'\!\!
   \int_0^{2\pi}\!\!\dd \theta'
   \frac{r' \sigma_i(r')}{\sqrt{r^2+r'^2-2rr'\cos\theta'}}.
   \label{eq:Vcontact}
\end{equation}
Here, $\tilde V_i$ are the contributions
to the surface potential from the surface charge
outside the contact region.
The integral are those from the surface charges
in the contact region.
The prefactor is $\frac 1{2\pi}$ (instead of $\frac 1{4\pi}$)
because of the well-known jump condition 
for surface potentials.\cite{kellogg1953foundations}
The distance at denominator is approximated using
that for the flat surface,
which leads to negligible error because
only the contact region is of concern.
The upper bound is set to infinity for convenience;
as we shall see below, the singular surface charge
density dies off rapidly outside the contact region.

The surface charge density~$\sigma_i(r)$
in eq.\,(\ref{eq:Vcontact}) are related to the
potential difference,
$\Delta V(r) \equiv \phi_1(r) - \phi_2(r)$.
By analogy to the conductor case,
the electric field is approximately vertical
and its magnitude is
$\Delta V(r) / (h + r^2/a)$.
Then applying Gauss's law,
we get the charge density on the top surface
\begin{equation}
\sigma_1(r) = \eout
\left(1 - \epsilon_{{\rm r},1}^{-1}
\right) \frac{\Delta V(r)}{h + r^2/a} ,
\label{eq:sigma1}
\end{equation}
where~$\epsilon_{{\rm r},1} \equiv \epsilon_1 / \eout$.
The charge density~$\sigma_2$
is given similarly, but with a negative sign.
Substituting the charge densities to eq.\,(\ref{eq:Vcontact})
and taking the difference
gives an integral equation for~$\Delta V(r)$. 
The dependence on the unknown, $\tilde{V}_1 - \tilde{V}_2$,
can be factored out by introducing
an auxiliary,
$ f(r) \equiv 1 - \Delta V(r)/(\tilde{V}_1 - \tilde{V}_2)$,
which satisfies
\begin{equation}
    f(r) =  
    \frac{\epsilon_{{\rm r},1}^{-1} + 
    \epsilon_{{\rm r}, 2}^{-1}}{2\pi}
    \int_0^\infty\!\!\dd r'\
    \frac{1-f(r')}{h+r'^2/a}
    \frac{4r'}{r+r'}
    K(x).
    \label{eq:fintegral}
\end{equation}
Here the integral over the azimuthal angle
has been replaced with the complete elliptic function
of the first kind $K(x)$,
where $x \equiv \frac{4rr'}{(r+r')^2}$. 
When~$\epsilon_1 = \epsilon_2$,
eq.\,(\ref{eq:fintegral}) reduces to Batchelor's original
result, eq.\,(4.5) in ref.\citenum{batchelor1977thermal}.
Since~$f(r)$ is proportional to the contribution
from the surface charges in the contact region,
we see that it is dominated by
particles of lower permittivity.
When both permittivities approach infinity,
we have $f(r) = 0$ and~$\Delta V(r) = \tilde{V}_1 - \tilde{V}_2$,
which is identical to the expression
for the conductors discussed above.
More general cases are discussed below.

\begin{figure}[t!]
    \centering
    \includegraphics[width=0.45\textwidth]{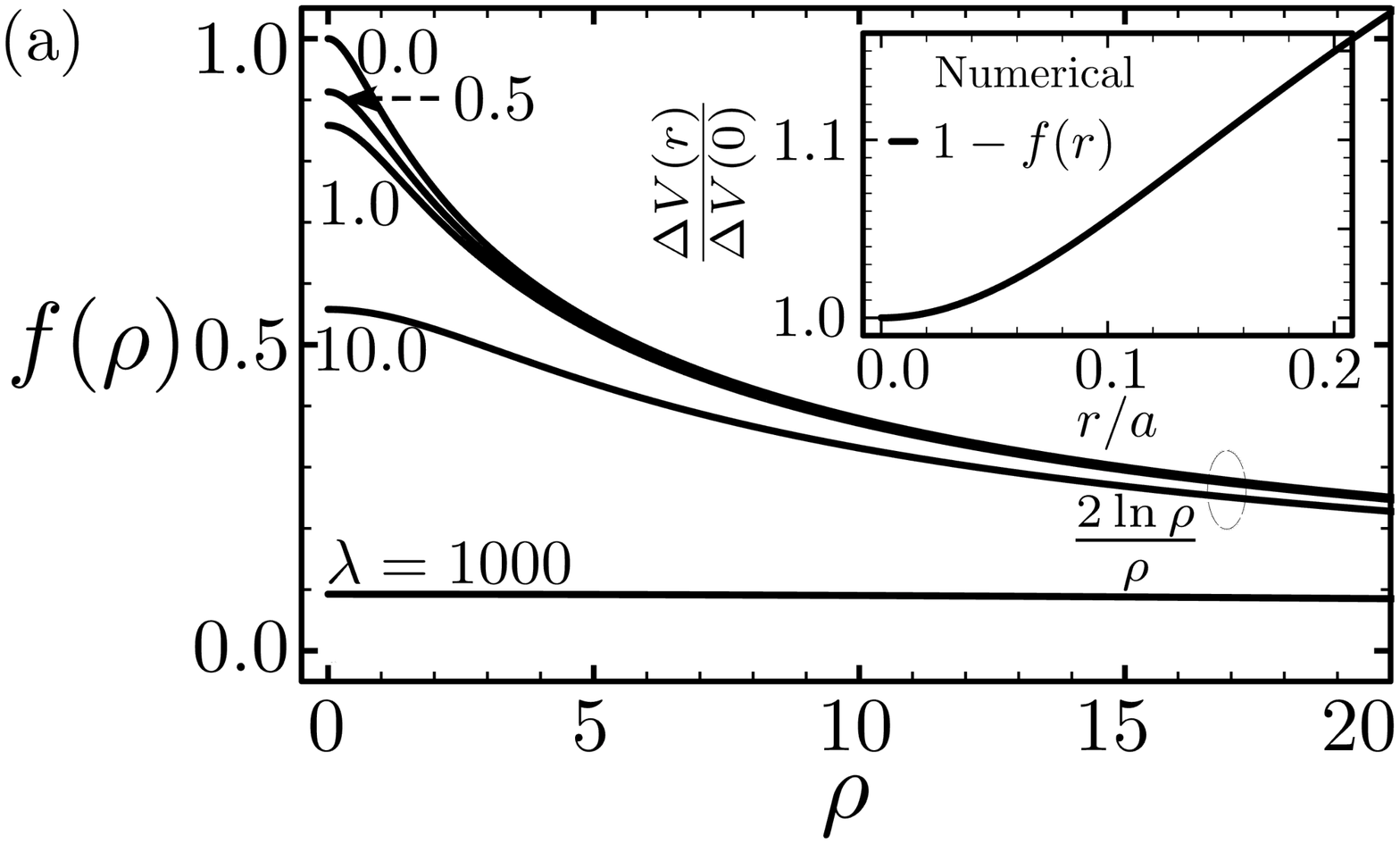}
    \includegraphics[width=0.45\textwidth]{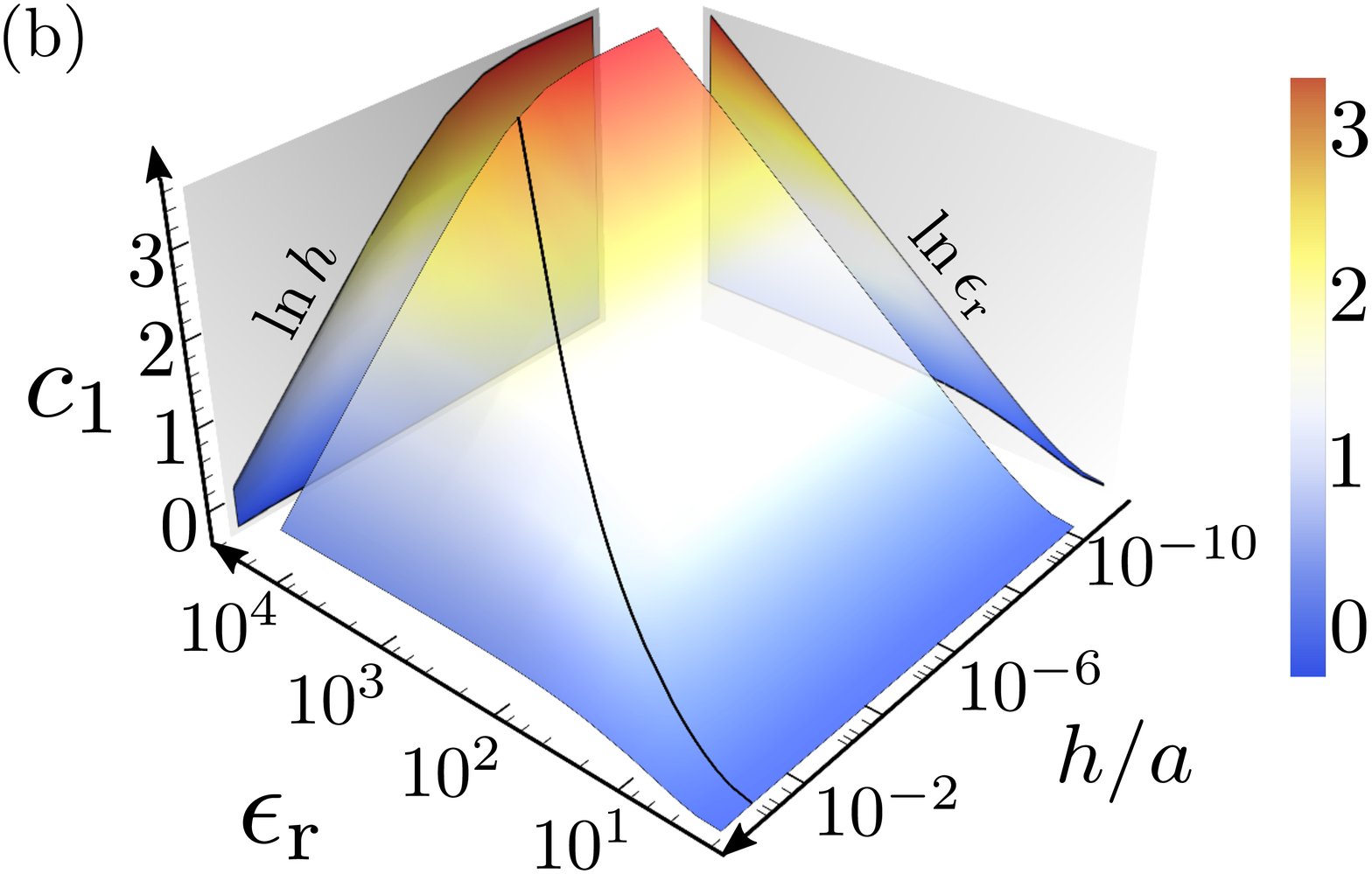}
    \caption{(a) Variation of surface potential~$f(\rho)$
    along the radial direction for $\lambda = 0$,
    $0.5$, $1.0$, $10.0$ and $1000$.
    The inset shows the normalized potential difference 
    for spherical dimer with a minimum separation $h=0.1a$ and parameters $(Q_i,\epsilon_i,a_i)$:
    $(1,100,1)$ and $(-1,10,2)$.
    Dots are numerically solved by a spectral method in bispherical coordinate \cite{lian2018polarization}
    and the solid line is proportional to $1-f(r)$. 
    (b) The singular capacitance $c_1$ is presented as a function of normalized surface separation $h/a$ and relative permittivity $\er$.
    }
    \label{fig:capacitance}
\end{figure}

The solution to eq.\,(\ref{eq:fintegral}) is
uniquely determined by the normalized distance $h/a$
and the average permittivity,
$\er\equiv2/(\epsilon_{\rm r,1}^{-1}+\epsilon_{\rm r,2}^{-1})$.
As shown by Batchelor,\cite{batchelor1977thermal}
eq.\,(\ref{eq:fintegral})
can be non-dimensionalized to
\begin{equation}
    f(\rho) = \frac{1}{\pi}\int_0^\infty\!\dd \rho'\,
    \frac{1-f(\rho')}{\lambda+\rho'^2}
    \frac{4\rho'}{\rho+\rho'}
    K(x)
    \label{eq:frho}
\end{equation}
in which $\rho\equiv r \er / a$
and $\lambda\equiv h \er^2/a$.
The numerically solved $f(\rho)$
for a few representative $\lambda$ values are 
shown in Fig.\,\ref{fig:capacitance}a.
For large gap distance, with $\lambda\gg1$,
$f(\rho)$ is nearly uniform, as expected.
For smaller $\lambda$ values,
$f(\rho)$ decreases from~$f(0)$ with~$r$ monotonically.
The difference in electric potential at the contact points
is proportional to $1 - f(0)$.
The value of~$f(0)$ increases as~$h$ decreases,
and reaches~unity at~$h = 0$,
ensuring that the surface potential is continuous
at the contact point.
The variation of~$f(\rho)$ obtained from the above local analysis
is confirmed by directly solving the full potential distribution
for dielectric dimers
(inset, Fig.\,\ref{fig:capacitance}a). 

Similar to eq.\,(\ref{eq:QsingularC}) for the conductor case,
the singular part of surface charges
on particle~$1$ is given,
with~$R_0$ being the regularizing cutoff of order~$a$, by
\begin{equation}
    Q_{\rm s,1} = \frac{(\tilde{V}_1 - \tilde{V}_2)\eout}{4\pi}
    \left(1 - \epsilon_{{\rm r},1}^{-1}\right)
    \int_0^{R_0}\!\!\dd r
    \frac{2\pi r \left[1 - f(r) \right]}{h+r^2/a}  
    \label{eq:QsingularD}    
\end{equation}
The dependence on the dielectric permittivity
appears in the prefactor~$1 - 1/\epsilon_{{\rm r},1}$ and
in~$f(r)$.
The singular surface charge~$Q_{\rm s,2}$ on the particle~2
is given analogously, but with a negative sign.
However, because of the dependence on the dielectric permittivity,
$Q_{{\rm s},2}$ and~$Q_{{\rm s},1}$ do not add up to zero.
The first term in the square bracket gives
the same~$\ln(a/h)$ singular contributions as eq.\,(\ref{eq:QsingularC}).
The second term represents the correction due to dielectric screening.

The singular capacitance defined
by~$c_1(h) \equiv Q_{{\rm s},1} / (\tilde{V}_1 - \tilde{V}_2)$
also has these two contributions.
In the non-dimensionalized form, it reads
\begin{equation}
    \begin{aligned}
    c_1(h) &=
    \frac{1 - \epsilon_{{\rm r},1}^{-1}}{4}a\eout
    \left(\ln \frac{a}{h} - P\left(\lambda\right)\right),\quad \\ 
    P(\lambda) &\equiv
    \int_0^{\infty}\!\dd \rho\, 
    \frac{2 \rho}{\lambda + \rho^2} f(\rho) .
     \end{aligned}
    \label{eq:singularC1}
\end{equation}
In the last term, the upper bound is
set to infinity
because~$f(\rho)$ decays rapidly
(as~$\ln \rho / \rho$, ref.\citenum{batchelor1977thermal}).
The dielectric correction is contained in the term~$P(\lambda)$,
which depends on permittivity~$\er$ and relative gap distance~$h/a$
through the combination~$\lambda = h \er^2 / a$.
The behavior of~$P(\lambda)$ is derived
from that of~$f(\rho)$.
For~$\lambda \gg 1$, $P(\lambda)$ is vanishingly small,
so the capacitance is dominated by~$\ln (a/h)$,
the characteristic behavior of conductors.
On the other hand, when~$\lambda \to 0$,
the leading contribution to $P(\lambda)$ is~$-\ln \lambda$,
which cancels exactly the~$\ln(a/h)$ dependence,
leaving a term~$2\ln\er$ that diverges instead
with the average permittivity.
Further, we note that~$c_1(h)$ approaches
the conductor limit for~$\er \gg 1$.

The difference in contact charges
between dielectric 
and conductor cases
is solely contained in~$P(\lambda)$.
For intermediate separation,
$c_1(h)$ shows a singular~$\ln(a/h)$ dependence
that is similar to the conductor behavior.
However, as long as~$\er$ is finite,
this logarithmic behavior will eventually be
cut off by contribution from~$P(\lambda)$
at sufficiently small separation.
Therefore, unlike the conductor case,
{\it the contact capacitance for dielectrics
is finite} at~$h = 0$.
Instead, it approaches
a constant proportional to~$2\ln\er$.
Physically, the logarithmic $h$-dependence
originates from the accumulated polarization charges at the interface.
It is cut off for dielectrics because the potential difference,
which gives rise to the polarization charge,
is self-consistently determined by the latter.
The weaker polarization effect for dielectrics
eventually can not keep up with the needed potential
difference for producing the
polarization charge.
The variation of~$c_1$ on~$h$ and~$\er$
are shown in Fig.\,\ref{fig:capacitance}b.
The crossover can be estimated by setting
$\lambda = h \er^2 / a = 1$.
The logarithmic divergence, although being cut at small gap separation,
still plagues numerical calculations in practice.\cite{jiang2016n}

This type of crossover behavior has been confirmed
in a study on dielectric spherical dimers using the image method.\cite{poladian1988general}
The limiting value of the capacitance~$2\ln \er$
as well as the crossover is also consistent
with the analytically known result
between a conducting sphere and a dielectric plane.\cite{kalinin2004nanoelectromechanics}
However, unlike these two work on dimer particles,
Batchelor's result is strictly local, suggesting 
that the same type of singularity
as represented by eq.\,(\ref{eq:singularC1})
is applicable to all contact region in
cluster of multiple dielectric particles,
irrespective of the particle shape.
In the following, we show that,
the approach for isolating the singularity for contact between conductors
can be generalized to treat the dielectric clusters,
yielding the exact contact energy and distance-dependence
with modest numerical cost.

We consider the cluster consisting of $n$~dielectric spheres.
Given the free charge distribution~$\rhof(\br)$,
the potential $\phi$ is governed by Poisson's equation
$\nabla\cdot\epsilon(\br)\nabla\phi =
-{\rhof(\br)}/{\eout}$,
where $\eout$ is the medium permittivity.
The (relative) material permittivity $\epsilon(\br)$
is set to $\er \equiv \ein / \eout$ inside particles and unity in the medium.
Although the general charge distribution poses no added difficulty,
for simplicity, we focus on the case when only
the free surface charge~$\sigma_{\rm f}$ are present.
In such case, the system energy can be written
as surface integrals of the product of the free surface charge
and the surface potential over all surfaces.
Furthermore, when the surface charge distribution is uniform,
the energy reduces to the sum of the product
of the total charge and the average surface potential.
Therefore, we denote the set of net charges on particles
by ${{\bf Q}}$,
and the mean surface potentials ${{\bf V}}$.
The net charges and the mean potentials are linearly related, i.e.,
\begin{equation}
    {\bf Q} = {\bf C} {\bf V},
    \label{eq:QCV}
\end{equation}
where ${\bf C}$ is an $n{\times}n$ ``capacitance array''.
In this notation, the total energy can
be expressed as
$\mathcal{E}=\frac{1}{2} {{\bf Q}}\cdot{\bf C}^{-1}\cdot{\bf Q}$.
For convenience,
the total energy $\mathcal{E}$ presented below 
is always normalized by the self-energy 
of a single isolated sphere with the radius $a$ and a net charge $q$,
$q^2/(8\pi\eout a)$.
We note that this formulation applies to arbitrary particle shapes
and cluster configurations.
If the surface charge distribution is non-uniform or any external excitation exists,
a multipole expansion of surface charge
in terms of dipole, quadrupole etc. is needed.
The constitutive relation eq.\,(\ref{eq:QCV})
and the quadratic expansion to energy can be generalized
to include the contribution from these higher multipoles 
and external electric fields.
For the purpose of demonstrating how the contact singularity
is isolated, we focus on the case of the uniform surface charge distribution. 

Our method for dielectrics is an extension 
to our earlier work on conductors.\cite{qin2016singular}
Because the conductors are equipotential,
the mean potential is also the surface potential 
at every point on the surface.
Since the capacitance~$\bf C$ in eq.\,(\ref{eq:QCV})
contains two types of contributions,
one type dominated by close contacts between neighboring particles,
and the other type from the remaining interactions from all particles,
we decompose the capacitance $\bf C$ as follows
\begin{equation}
    {\bf Q} = \left(c_{\rm s}{\bf L} + {\bf H}\right)
    {\bf V}. 
    \label{eq:CdecompositionC}
\end{equation}
Here, $c_{\rm s}(h) = \frac{a\eout}{4}\ln(a/h)$ 
is the singular capacitance for conductors derived above
(assuming that all gap distances are~$h$).
Since~$c_{\rm s}$ becomes significant only for small gaps,
the entries in the array~$\bf L$ are nonvanishing only for
closely neighboring particles.
Specifically, if particle~$i$ and~$j$ form a close contact,
we set the entries $L_{ij} = L_{ji} = -1$.
The diagonal entry $L_{ii}$ equals
the number of close neighbors of the particle~$i$.
All other entries of~$\bf L$ are set to zero.
The singular capacitance~$\bf L$ as defined here is
the same as the adjacency matrix
representing the connectivity of clusters
(see ref.\citenum{qin2016singular} for explicit examples).
In practice, for a given cluster configuration,
we first numerically solve
the full capacitance array~$\bf C$
at finite but small~$h$ values,
then subtract from~$\bf C$
the singular term~$c_{\rm s}(h) {\bf L}$,
to get the regular capacitance~${\bf H}$.
The $h$-dependence of this regular capacitance
is then fitted to a straight line when~$h$ is small,
allowing us to obtain
the full $h$-dependence for the capacitance array~$\bf C$
and, consequently,
the full $h$-dependence of energy, down to~$h = 0$.

Generalizing eq.\,(\ref{eq:CdecompositionC}) to dielectrics
requires two nontrivial modifications.
First, the decomposition in eq.\,(\ref{eq:CdecompositionC})
is valid because the potential of the conductors
can be used to evaluate the contact potential difference.
But dielectrics are not equipotential,
and the surface potential at the contact points~$\tilde V_i$
generally differ from the average potential~$V_i$
by a numerical factor that depends on
the dielectric permittivity and the cluster configuration.
Its variation with gap distance~$h$ is weak,
and approaches a constant as~$h \to 0$.
So we generalize eq.\,(\ref{eq:CdecompositionC}) to
\begin{equation}
    {\bf Q}=\left(\tldL + {\bf H}\right){\bf V},\quad
    \tilde{L}_{ij} \equiv
    \frac{1 - \epsilon_{{\rm r},j}^{-1}}{V_j}
    \begin{cases}
    \tilde V_j^{(i)} c_{ij},
    & i \neq j ; \\[5pt]
    {\sum_{k=1}^N} \tilde V_j^{(k)} c_{jk} ,
    & i = j .
    \end{cases}
    \label{eq:capdielectric}
\end{equation}
Above, the superscript `$(i)$' in~$\tilde V_j^{(i)}$
indicates that the contact potential is evaluated
on the particle~$j$ at the contact formed with the particle~$i$.
The singular capacitance~$c_{ij}$ 
is given from eq.\,(\ref{eq:singularC1}) by
$c_{ij} = \frac{\eout a_{ij}}{4}
\left(\ln \frac {a_{ij}}{h_{ij}}
- P(\lambda_{ij})\right)$,
which depends on the mean radius of curvature~$a$,
gap distance~$h$,
and~$\lambda$ value evaluated for the particle pair~$i$ and~$j$.
From the definition, it is clear that
$\tldL$ is generally not symmetric,
which however reduces to the (symmetric) form
identical to eq.\,(\ref{eq:CdecompositionC})
in the conductor limit,
since~$\tilde V_j^{(i)} = V_j$ for all contacts on the particle~$j$.

Second, it turns out that the contact potentials
exhibit strong dependence on the second
order and more distant neighbors,
because the dielectric screening is comparatively weak 
(see Fig.\,\ref{fig:cubeE} below).
Therefore, to ensure rapid convergence
of the correct contact potentials,
our construction of the adjacency array~$\tldL$
contains all pairs of particles.
Fortunately, as we shall see below,
this construction requires no extra computation
other than evaluating the surface potentials
at additional contact points.

In order to identify the regular part ${\bf H}$
using eq.\,(\ref{eq:capdielectric}),
we numerically compute the full capacitance matrix ${\bf C}(h)$
and all the contact potentials~$\tilde V_j^{(i)}$,
for a range of small but nonzero $h$ values.
We used the boundary element method (BEM) 
implemented in the package COPSS \cite{jiang2016n}
to solve Poisson's equation.
The solution is expressed in terms of
the induced bound (polarization) charges $\sigma_{\rm b}$,
which satisfies the following boundary integral equation,
\begin{equation}
    \frac{1 + \er}{2}\sigma_{\rm b} 
    + (1 - \er)\eout \left( {\bf E}_{\rm b} + {\bf E}_{\rm f}\right)
    \cdot\hatbn
    = \frac{1 - \er}{2}
    \sigma_{\rm f} . 
    \label{eq:BEM}
\end{equation}
Above, the electric field strength at
the surface due to induced bound charge~$\sigma_{\rm b}$
and free charges~$\sigma_{\rm f}$ are
expressed as surface integrals
($\alpha = {\rm b, f}$),
\begin{equation}
    {\bf E}_{\alpha} = \frac{1}{4\pi \eout}
    \int\! \dd S' \,\frac{\br-\br'}{|\br-\br'|^3}\sigma_{\alpha}(\br').
    \label{eq:Esurface}
\end{equation}
By our convention, the surface normal~$\hatbn$ points outward.
For clusters of multiple particles,
it is understood that different particles
may have different values of permittivity~$\epsilon_{\rm r}$.
Equation~(\ref{eq:BEM}) is linear in~$\sigma_{\rm b}$
and~$\sigma_{\rm f}$.
After discretization, the bound charge is obtained
by inverting the coefficient array.
The apparent divergent diagonal entries
while discretizing the surface integral~eq.\,(\ref{eq:Esurface})
can be re-parameterized to
reproduce the correct self-energy.
Further numerical details can be found in ref.\citenum{barros2014efficient}. 
In all calculations by BEM presented in this work
(the dimer example used the
exact series expansion.\cite{lian2018polarization}),
we meshed each spherical surface into $17342$ triangular patches
such that mesh size is about $0.03\,a$.

In general, for a cluster of $n$ particles,
$n$ separate numerical calculations with
$n$ independent charge vectors $\bf{Q}$ 
are needed to determine the full capacitance matrix ${\bf C}$. 
The computational cost can be reduced
by imposing the symmetry of the cluster configuration.
Subtracting from~$\bf C$ the singular contribution~$\tldL$
is expected to give the $h$-dependence of the regular part~$\bf H$.
However, unlike the conductor case,
where the coefficient to the logarithmic term is exactly known,
the singular capacitance $\tldL$ for dielectrics
depends on the calculated
contact potentials $\tilde{V}_j^{(i)}$,
which is susceptible to the numerical precision
and what is meant by `contact point' on a surface mesh.
In practice, we vary the gap distance $h$
and compute a few trial values of contact potentials,
then select the one that ensures the difference
$H_{ij}(h) = C_{ij} - \tilde{L}_{ij}$ converges
linearly with~$h$ as $h\to0$ for all the entries. 
In the following, three examples are presented
to demonstrate the nature of the contact singularity,
and to show that the regularization scheme allows us to
obtain the energy of cluster of dielectric particles
with small $h$.

The first example is a dimer of identical dielectric spheres.
It is the simplest example for demonstrating the effects
of interfacial polarization.\cite{russell1909coefficients,lindgren2018integral,lian2018polarization,love1975dielectric}
Except a study on the interaction between a conducting sphere
and a dielectric plane,\cite{kalinin2004nanoelectromechanics}
very few past work tried to evaluate the energy at small separation,
presumably due to the difficulty of resolving the aforementioned contact singularity. 
Therefore, we analyzed the polarization effect
for dielectric dimers in close contact,
verified the singular behavior in eq.\,(\ref{eq:singularC1}),
and showed that eq.\,(\ref{eq:capdielectric})
yields the full $h$-dependence of energy.
We studied both symmetric (${\bf Q} = (q, q)$) and
asymmetric (${\bf Q} = (q, -q)$) cases,
for a range of permittivity values ($1 \leq \er \leq 10^6$).
The energy at~$h = 0$ is presented as a function
of relative permittivity, which agrees
with the exact result (see SI) found
using the tangent-sphere coordinate.

\begin{figure}[t!]
    \centering
    \includegraphics[width=0.5\textwidth]{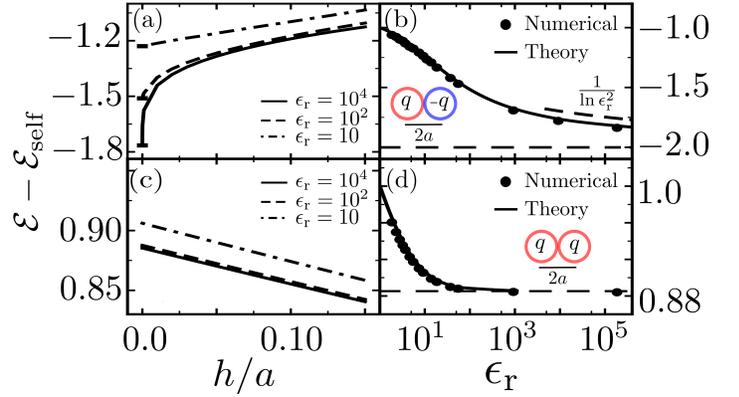}
    \caption{Electrostatic energy of a pair of
    identical spheres.
    (a) and (b): asymmetric charge ${\bf Q} = (q, -q)$.
    (c) and (d): symmetric charge ${\bf Q} = (q, q)$.
    In (b) and (d), dots are
    numerical results obtained using our proposed method,
    and curves are exact prediction obtained using
    the tangent-sphere coordinate (SI).
    }
    \label{fig:dimerE}
\end{figure}

For the asymmetric case,
interfacial polarization enhances the inter-particle attraction.
To illustrate this effect, we subtract the energy of two
isolated spheres from that of a dimer,
then plot it against the gap distance in Fig.\,\ref{fig:dimerE},
for three representative values of $\er$:
$10$, $10^2$ and $10^4$. 
In absence of the polarization effect,
the energy curves would exhibit no dependence on~$\er$,
and appear flat over this narrow range of~$h/a$ values.
As~$\er$ is increased from~$1$,
we confirm the expected, stronger distance dependence,
as~$h \to 0$.
In particular, for the case of $\er=10^4$,
an apparent logarithmic dependence on $h$ 
is seen in Fig.\,\ref{fig:dimerE}a,
which is precisely the singularity revealed in eq.\,(\ref{eq:singularC1})
and is stronger for higher permittivity values.
On the other hand,
when the permittivity $\er$ is decreased, 
this logarithmic dependence is cut off at
an increasingly larger separation and
becomes almost invisible when $\er=10$.

The normalized energy difference, as shown in Fig.\,\ref{fig:dimerE}a,
evaluated at~$h = 0$, is show in Fig.\,\ref{fig:dimerE}b
for different~$\er$ values.
This contact energy represents the energy gain
for bringing two particles from infinity to contact.
In terms of the normalizing energy unit,
$q^2/(8\pi\eout a)$,
its value is~$-1$ without the polarization effect,
and becomes more negative as~$\er$ increases,
eventually reaching~$-2$ at~$\er = \infty$.\cite{qin2016singular}
The contribution from the polarization effect
for~$\er \gtrsim 100$ is comparable to the coulombic attraction alone.
Moreover, we notice a slow convergence
of contact energy for permittivities~$\er \gtrsim 10^3$,
owing to the weak $\ln\er^2$ dependence
in the contact capacitance eq.\,(\ref{eq:singularC1}) at~$h = 0$.
Finally, we affirm our results obtained from eq.\,(\ref{eq:capdielectric}),
by comparing the contact energies to
the exact analytical results (SI).

For the symmetric case, the interfacial polarization
weakens the inter-particle repulsion,
which is seen from the $h$-dependence of dimer energy.
However, by symmetry,
the potential difference in the gap region vanishes
and thus no logarithmic behavior is observed
in Fig.\,\ref{fig:dimerE}c.
The energy scales linearly with~$h$
and a straightforward extrapolation gives
the contact energy for all permittivity values
plotted in Fig.\,\ref{fig:dimerE}d.
A much weaker polarization effect is found.
Even for the conductor case,
only about $10\%$ contact energy can be attributed to polarization effects.
The contact energies converge for~$\er \gtrsim 100$,
and reach $2(1/\ln2-1)\simeq0.88$,
in agreement with Maxwell's result\cite{maxwell1873treatise} for the conducting dimers.
As for the asymmetric case,
the full $\er$-dependence matches
the analytical calculation (SI).
To conclude this example, we note that
the symmetric case is a peculiar example
where the contact singularity is strictly absent.
For all other charge ratios, using the decomposition
in eq.~(\ref{eq:capdielectric}) to isolate the strong
$h$-dependence is necessary.

The second example concerns the energy
of $8$~identical spheres placed at the vertices of a cube,
which illustrates the importance of the second order contact
for dielectric clusters.
As discussed above, the singular capacitance $\tldL$ in~eq.\,(\ref{eq:capdielectric})
contains not only contributions from the nearest neighbors,
but also those from the second-order and higher order neighbors.
Therefore, a sphere at a vertex of a cube
forms a secondary contact with its $3$~plane diagonal vertices,
and a third order contact with the body diagonal vertex.
Even though the higher order contributions are minor,
because at such large separation, these contacts cease to be singular,
keeping the second order contribution is essential
for obtaining the correct linear scaling of the regular capacitance
with gap distance~$h$ shown in Fig.\,\ref{fig:cubeE}a.

As in the dimer case, these regularized capacitance allow us
to evaluate the energy at arbitrarily small gap distance.
Figure~\ref{fig:cubeE}b shows the normalized energies
when only one particle is charged,
for which all the energetic contributions come from the interfacial polarization.
Comparing the results from BEM and our regularization schemes
containing varying order of contact contributions,
it is clear that keeping the higher order singular contribution is
necessary for correctly evaluating the energy for~$h/a \lesssim 0.05$.
In contrast, no such terms are needed for the conducting case.

\begin{figure}[t!]
    \centering
    \includegraphics[width=0.49\textwidth]{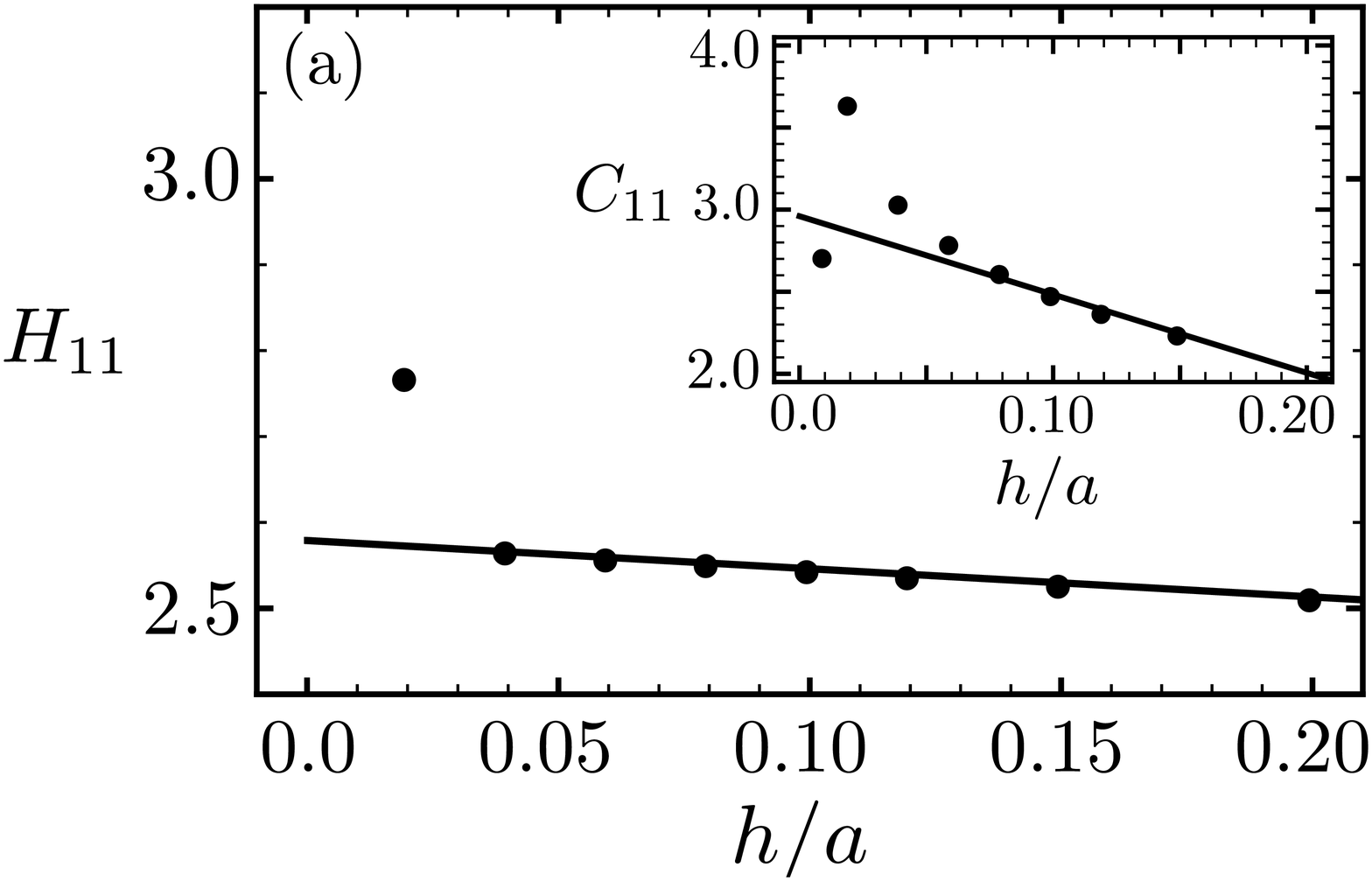}
    \includegraphics[width=0.545\textwidth]{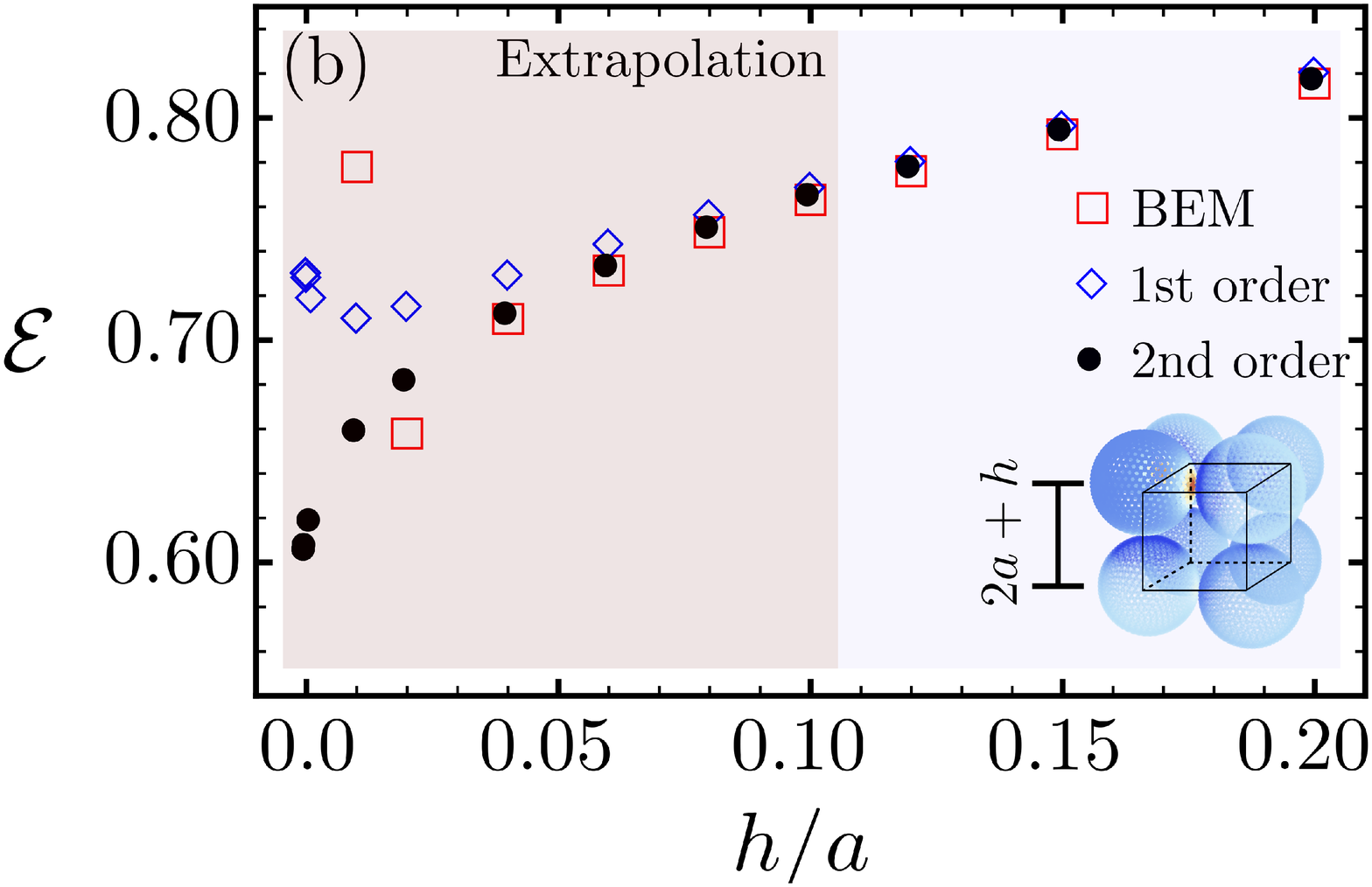}
    \caption{(a) Variation of entry~$H_{11}$ in the regular
    capacitance array against separation.
    The results for the other entries are
    provided in Fig.\,S3.
    (b) Variation of electrostatic energy against separation
    for dielectric spheres with $\er=100$ placed
    on cubic vertices, where one sphere is charged.}
    \label{fig:cubeE}
\end{figure}

The third example is our main result
on the energy of clusters,
from which the cohesive energy is obtained
by subtracting the self-energy of particles
at infinite separation.
We considered two limiting configurations:
the most extended one with all spheres arranged along a straight line (string),
and the most compact one with all spheres posited at the vertices
of the platonic solids (polyhedron).
In our earlier work on the conducting spheres,
the charges are allowed to flow freely between contacting spheres.
The energy of polyhedron packing is found to be lower than
that of the string packing at a finite separation.
However, as the gap distance decreases,
the polarization effects due to contact singularity become increasingly relevant,
which ultimately makes the string packing
to be energetically more favorable
than the polyhedron packing.
For the dielectric cases,
we show that the weakened contact singularity
causes another crossover between
the relative stability of string-like
and polyhedral packings.

\begin{figure}[t!]
    \centering
    \includegraphics[width=0.57\textwidth]{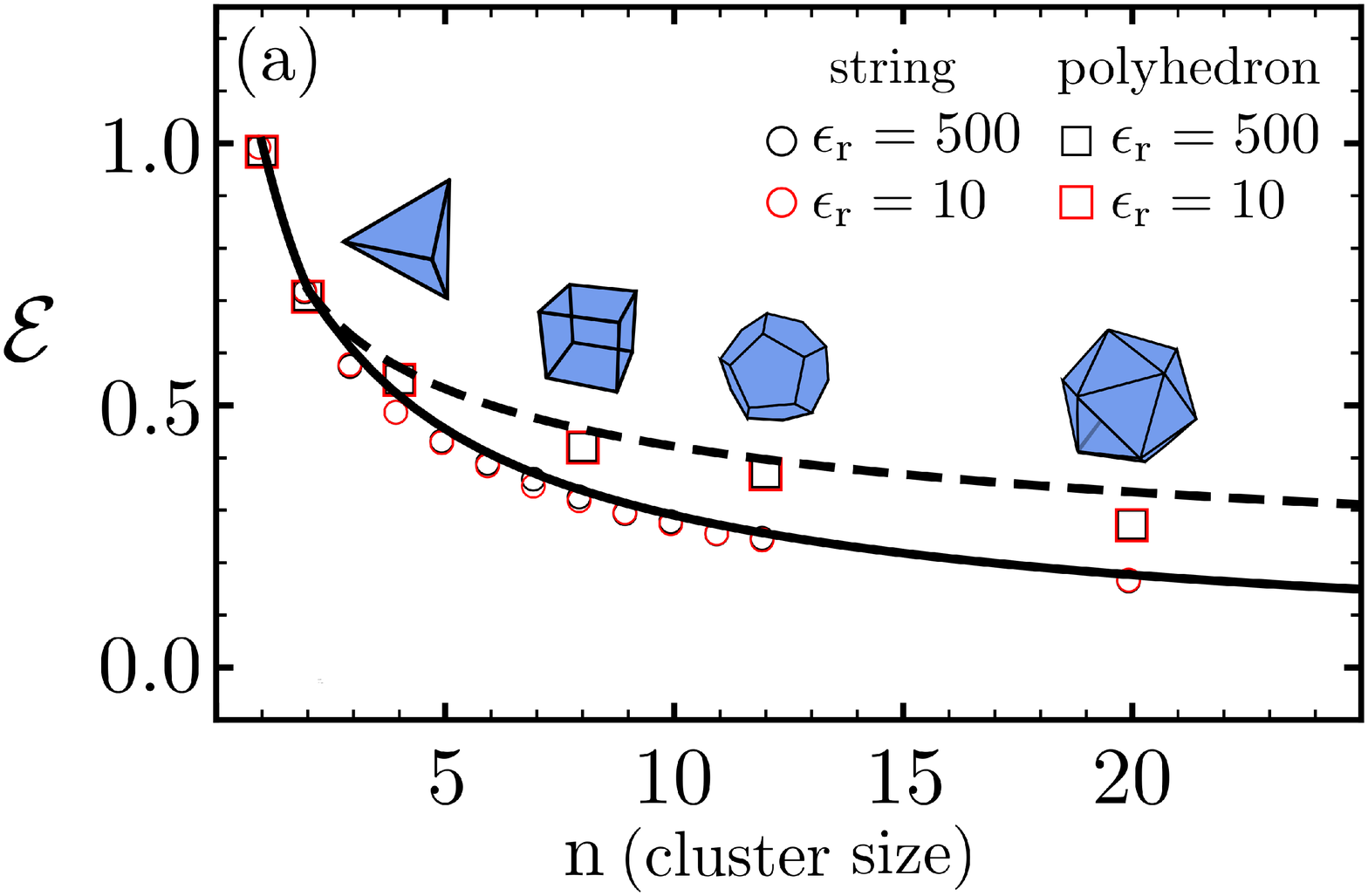}
    \includegraphics[width=0.49\textwidth]{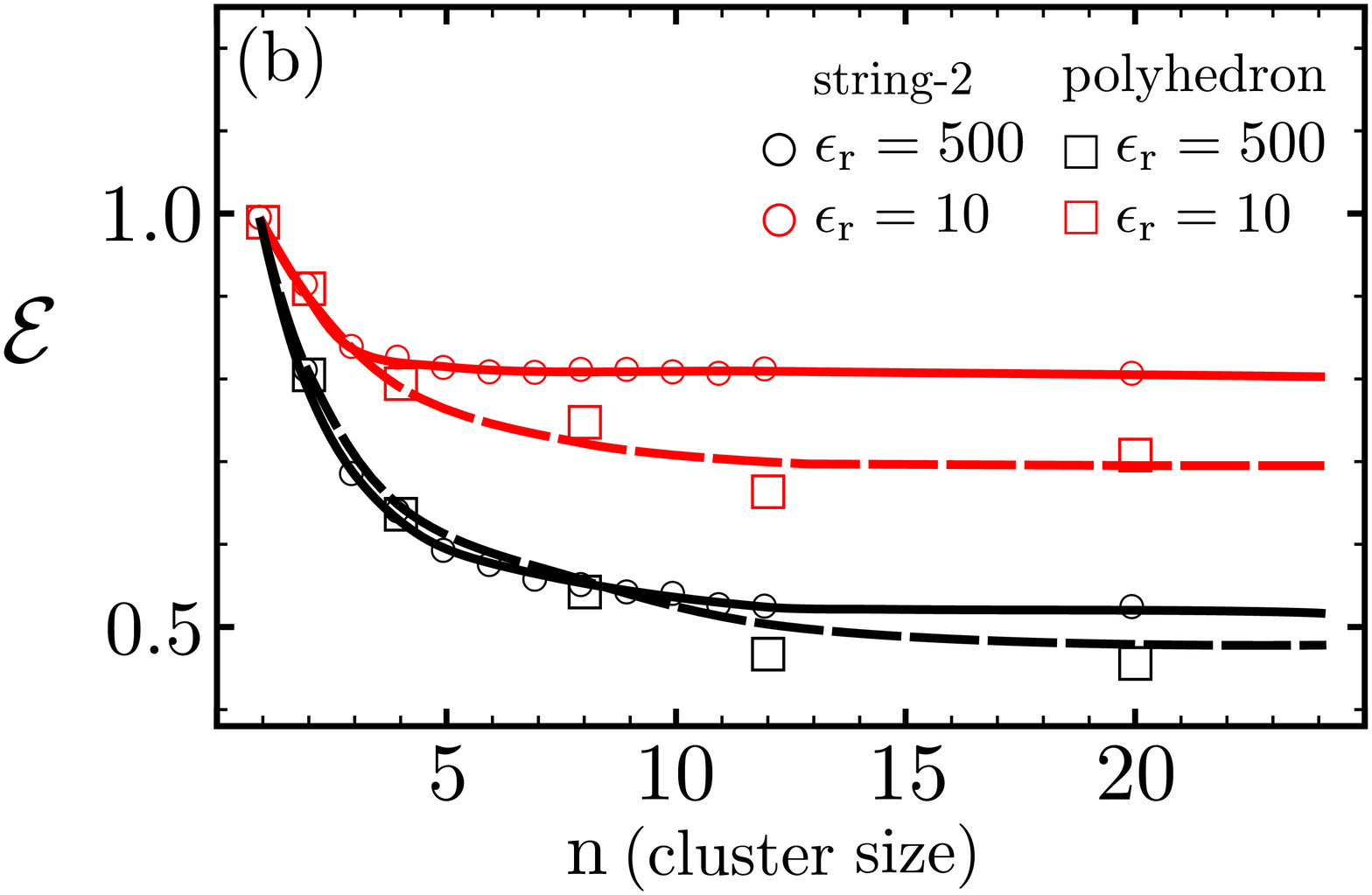}
    \caption{Size dependence of electrostatic energy $\mathcal{E}$ 
    for string-like and polyhedral configurations,
    with fixed total charge.
    (a) Charge transfer is permitted.
    Dashed curve: 
    $\mathcal{E}=(2/n)^{1/3}/(2\ln2)$,
    the energy for polyhedral configurations of conducting spheres.\cite{qin2016singular}
    Solid curve: $\mathcal{E}=\ln(2na/r_0)/[n\ln2\,\ln(2a/r_0)]$, the energy for a conducting cylinder
    of the same volume with length $L=2na$ and radius
    $r_0=0.816\,a$.\cite{jackson1999classical}
    (b) Charge transfer is prohibited.  
    The total charge $q$ resides on 
    the middle particle of the string (string-$2$)
    and an arbitrary vertex of polyhedron (polyhedron).
    }
    \label{fig:cluster}
\end{figure}

As the previous example, to highlight the polarization effects,
a unit charge~$q$ is placed on one sphere.
For symmetric polyhedron packing, this charged sphere is arbitrarily chosen.
For the string packing, we considered
two extreme placings:
at one end (string-1) and in the middle (string-2).
For each particle configuration and charge assignment,
we considered two scenarios:
charge transfer between contacting particles
is permitted (Fig.\,\ref{fig:cluster}a) and prohibited (Fig.\,\ref{fig:cluster}b). 

When inter-particle charge transfer is permitted,
while maintaining uniform charge distribution on each individual particle,
the energy $\mathcal{E}$ can be calculated using ${\bf Q}\cdot {\bf C}^{-1} \cdot {\bf Q} / 2$.
Minimizing this energy subject to the constraint
of constant total charge~$q$,
the optimal charge assignment is found to
be~${\bf Q}_e = q\, {\bf C} \cdot {\bf u} / C_{\rm e}$,
where~${\bf u} \equiv (1, 1, \cdots, 1)$
and~$C_{\rm e} \equiv \sum_{i,j=1}^n C_{ij}$.
This implies that the optimal charge~${\bf Q}_e$
produces identical {\it average} surface potentials
for all the particles,
which generalizes the `equipotential'
concept for conductors.\cite{qin2016singular}
Correspondingly, the minimized energy
is given by~$q^2 / (2 C_{\rm e})$,
where the mean capacitance~$C_{\rm e}$
weakly depends on the dielectric permittivity.
In this scenario, there is no need to
differentiate string-$1$ and string-$2$ configurations.
The energies of string and polyhedron configurations at~$h = 0$
are plotted in Fig.\,\ref{fig:cluster}a,
for ~$\er = 10$ and~$500$.
For all configurations,
the energy decreases with the cluster size $n$ monotonically
since the redistributed surface charges are further apart.
The dependence on permittivity is surprisingly weak,
and the size-dependence is nearly indistinguishable
from that of the conductor case.\cite{qin2016singular}
Two curves are obtained
by treating the cluster as a single conductor respectively,
whose capacitance scales with
the length scale of the cluster,
i.e., $n^{1/3}$ for the polyhedron packing
and~$n/\ln(n)$ for the string packing.\cite{qin2016singular}
The extended string packing has a lower cohesive energy
because its effective capacitance is higher
than that of the compact polyhedron packing.

More interesting behavior is found (Fig.\,\ref{fig:cluster}b)
when charge transfer is prohibited.
The dependence on permittivity is evident
for all three cases: string-$1$, string-$2$, and polyhedron.
To better visualize energies of string-$2$ and polyhedron packing, 
energies of string-$1$ is presented 
in Fig.\,S4)
The energy is lower for the cluster with higher permittivity,
because the polarization effect is stronger.
The energy of string-$2$ is lower than that of string-$1$
for identical~$n$,
because the charged particles in the middle of the string
can polarize the particles in two half strings.
Further, the energies for both string-$1$ and string-$2$ configurations
saturate as the cluster size grows beyond~$n = 12$,
because the polarization effect is short-ranged.
To assess the relative stability of
compact or extended configurations,
we then only need to focus on the string-$2$ and the polyhedron cases.

Our results indicate that the relative energetic stability
depends on the permittivity.
For $\er=10$, the energy of the polyhedron packing  
is lower than the string-2 packing,
which is opposite to
the characteristic result for conductors
shown in Fig.\,\ref{fig:cluster}a.
Therefore, we expect a crossover
from stable compact packing to stable open packing
at intermediate permittivity values.
This is indeed the case shown for~$\er = 500$,
whereby the energies of compact and open configurations
closely trace each other,
and the relative energetic stability changes at~$n=4$ and~$n=8$ respectively.
As~$\er$ is further increased,
the energy curves of corresponding configuration 
will converge to those in Fig.\,\ref{fig:cluster}a,
reversing the relative stability
of open and compact packings.

In summary, we generalized our previous work on conducting particles,\cite{qin2016singular}
and developed a scheme to resolve the singular contact charges
between touching dielectric spheres,
which regularizes the full capacitance array
by isolating the singular contributions,
i.e., eq.\,(\ref{eq:capdielectric}).
Using this scheme, we obtain
the cohesive energies for dielectric clusters at zero separation
containing up to~$n = 20$ particles,
which is difficult to resolve with brutal force numerical calculations.
Our results show that the shape of stable clusters
formed from dielectric particles depends
on the permittivity ratio $\er$:
open clusters is more stable for large~$\er$,
and compact clusters is more stable
for small~$\er$.
Our scheme is applicable to systems with
arbitrary packing geometry,
charge distribution, 
and containing asymmetric interfaces that have different permittivities and
radii of curvature on the contacting particles.
At the heart of our scheme is
the contact singularity,
which resembles those encountered 
in the study of thermal conduction
\cite{batchelor1977thermal}
and momentum transport across a narrow gap.\cite{davis1989lubrication}


\bibliographystyle{unsrt}
\bibliography{ms}

\end{document}


\maketitle


We provide a detailed derivation of 
a spectral method for the electrostatic problem 
of two touching dielectric spheres carrying uniform free surface charges 
in the tangent-sphere coordinates.
The similar problem of touching dielectric spheres 
in presence of an electrostatic field 
has been solved by Pitkonen previously.\cite{pitkonen2008polarizability} 
We begin by introducing the tangent-sphere coordinates and 
review some useful mathematical facts 
about the Poisson's equation in the tangent-sphere coordinates.
We then provide a step-by-step derivation of the theory 
and illustrations of the method at the end.

First, we introduce the notation of the 
tangent-sphere coordinates
by defining the transformation from the tangent-sphere coordinates $(\mu,\nu,\varphi)$
to the Cartesian coordinates $(x,y,z)$
\begin{equation}
    x=\frac{\mu\cos\varphi}{\mu^2+\nu^2}, \quad 
    y=\frac{\mu\sin\varphi}{\mu^2+\nu^2}, \quad 
    z=\frac{\nu}{\mu^2+\nu^2}.
    \label{eq:cyltotan}
\end{equation}
where $z$-axis is the line connecting centers of two spheres. 
In this curvilinear coordinate,\cite{moon2012field}
spherical surfaces tangent to $xy$-plane at the origin
have constant $\nu$ value in $\nu\in(-\infty,+\infty)$.
The value of $\nu$ is determined by the radius of the spherical surface 
through $\nu=\pm 1/(2a)$,
where $+$ and $-$ indicate the surfaces with $z>0$ and $z<0$, respectively.
$\nu=0$ denotes the $xy$-plane thereby. 
The surface of a circular toroid centered at origin without a hole, 
whose radius of circular section is $\mu$, 
has a constant $\mu$ value in $\mu\in[0,+\infty)$.
Lastly, $\varphi\in[0,2\pi]$ is the azimuth angle. 
The Euclidean distance $d$ between points $(\mu,\nu,\varphi)$ and $(\mu',\nu',\varphi')$ is expressed as 
\begin{equation}
      d= \left( \frac{\mu^2-2\mu\mu'\cos(\varphi-\varphi')+\mu'^2
      +(\nu-\nu')^2}{(\mu^2+\nu^2)(\mu'^2+\nu'^2)}\right)^{1/2}
\end{equation}

The electrostatic potential $\phi$ is governed by 
the Poisson's equation $\nabla\cdot\epsilon(\br)\nabla\phi(\br)=-\rho_{\rm f}(\br)/\epsilon_0$,
where $\epsilon(\br)$ and $\epsilon_0$ are  
the relative permittivity and the vacuum permittivity,
and $\rho_{\rm f}(\br)$ is the free charge distribution. 
$\rho_{\rm f}(\br)$ is nonzero only if $\br$ is on the surfaces in our problem setting. 
We denote $\phi_i (i=1,2)$ as the potential inside the sphere $i$ 
and $\phi_0$ as the potential outside two spheres.
Within each region, the potential $\phi_i$ satisfies 
the Laplace's equation $\nabla^2\phi_i=0$
and then the solution $\phi_i$ can be deduced from 
the general solution of Laplace's equation 
in tangent-sphere coordinates by matching the values of $\phi_i$ on boundaries.

The potential function $\phi(\mu,\nu,\varphi)$ 
satisfying the Laplace's equation 
in the tangent-sphere coordinates
is $R-$separable \cite{moon2012field},
in which $R(\mu,\nu)\equiv(\mu^2+\nu^2)^{1/2}$, 
and has the following general form 
\begin{equation}
    \begin{aligned}
    \phi(\mu,&\nu,\varphi)= R(\mu,\nu) 
    \int_0^\infty\!\dd \lambda\, \lambda J_n(\lambda\mu)
    \left(A_n(\lambda) e^{\lambda\nu}+B_n(\lambda) e^{-\lambda\nu}\right)
    \left\{\begin{aligned}
        \cos n\varphi \\ 
        \sin n\varphi
    \end{aligned}\right\}.
    \end{aligned}
    \label{eqA:general}
\end{equation}
Above, $n$ takes nonnegative integrers $n=0,1,2,\ldots$,
$J_n(x)$ is the $n$-th order Bessel function of the first kind. 
$A_n(\lambda), B_n(\lambda)$ are two continuous spectra of $\lambda$ 
that will be determined through matching boundary values.
The integral above is the Hankel transform of order $n$. 
The definition of Hankel transform of order $n$ 
and its inverse transform are 
\begin{subequations}
   \begin{align}
   \Phi_n(\mu,\nu) &= \int_0^\infty\,\dd\lambda\, \lambda J_n(\lambda\mu)\overline{\Phi_n}(\lambda,\nu), \\ 
   \overline{\Phi_n}(\lambda,\nu)&=\int_0^\infty\,
   \dd\mu\,\mu J_n(\lambda\mu)\Phi_n(\mu,\nu) . 
   \end{align}
\end{subequations}
Owing to the rotational symmetry of the dimer problem we concerned, 
we shall only need the solution with $n=0$ and 
thus the $\varphi$-dependence can be dropped 
in our discussion below. 
We refer the further details about tangent-sphere coordinates 
to ref.\,\cite{moon2012field}

We present below a step-by-step derivation of the spectral method  
for the case of asymmetric free charges ${\bf Q} = (q, -q)$.
The other case of symmetric free charge ${\bf Q} = (q, q)$ is dealt 
in the identical procedure, 
whose final form of solution would be provided directly 
in parallel with that of the asymmetric case. 
To simplify the algebra, 
we consider the dimer of identical spheres 
with radius $a_1=a_2=a=\frac{1}{2}$ 
and define the permittivity contrast $\epsilon_{\rm r}=\ein/\eout$, 
where $\ein$ and $\eout$ are the relative permittivities of the particles and the medium. 
The vacuum permittivity $\epsilon_0$ is set to the unity in below.
The surfaces of two spheres then have the constant value $\nu=\pm1$ in our convention
and the centers of sphere are $(0,\pm2,0)$ in the tangent-sphere coordinates. 
For convenience, the potentials $\phi$ and energies $\mathcal{E}$ shown below 
are always normalized by 
the self potential and self energy of a single isolated sphere in the vacuum,  
i.e. $q/(4\pi \epsilon_0 a)$ and $q^2/(8\pi \epsilon_0 a)$, respectively. 

For the dimer with asymmetric free charges, 
the potential has to be antisymmetric 
with respect to the $xy$-plane,
i.e. $\phi(\mu,\nu) = -\phi(\mu,-\nu)$. 
Therefore, we only need to consider the potential $\phi_1$ and $\phi_0$
in the upper half space $\nu>0$.
The potentials outside the sphere $\phi_0$
and inside the sphere $\phi_1$ are 
\begin{equation}
    \begin{aligned}\phi_0(\mu,\nu) &=
     \frac{1}{\eout}\left(\frac{(\mu^2+\nu^2)^{1/2}}{(\mu^2+(\nu-2)^2)^{1/2}}
     -\frac{(\mu^2+\nu^2)^{1/2}}{(\mu^2+(\nu+2)^2)^{1/2}}\right) 
       +\frac{1}{\eout}\psi_0(\mu,\nu),  
    \\
    \phi_1(\mu,\nu) &=
    \frac{1}{\eout}\left(1
    -\frac{(\mu^2+\nu^2)^{1/2}}{(\mu^2+(\nu+2)^2)^{1/2}}\right) 
    +\frac{1}{\eout}\psi_1(\mu,\nu). 
    \end{aligned}
    \label{eqA:potential}
\end{equation}
Here, the first part of contribution in the parenthesises come from the free charges of two spheres.
The second part, potentials $\psi_0$ and $\psi_1$, 
represents the contribution from the induced bound charges. 
One can show that the first part satisfies the Laplace's equation by themselves alone,
which follows that the $\psi_0$ and $\psi_1$ do the same. 

The potentials $\psi_0$ and $\psi_1$ generated by the induced bound charges 
can then be expressed in the form of eq.\,(\ref{eqA:general}) 
\begin{subequations}
    \begin{align}
        \psi_0(\mu,\nu) &=R(\mu,\nu) 
        \int_0^\infty\!\dd \lambda\,
        \lambda J_0(\lambda\mu)
        \left(
        2A^{(0)}(\lambda) \sinh{\lambda\nu}
        \right),       
        \label{eqA:psi0}
        \\ 
        \psi_1(\mu,\nu) &= 
        R(\mu,\nu)
        \int_0^\infty\!\dd \lambda\,
        \lambda J_0(\lambda\mu)
        \left(
        B^{(1)}(\lambda) e^{-\lambda\nu}\right).       
        \label{eqA:psi1}
    \end{align}
\end{subequations}
The superscript `$(i)$' of the spectra 
$A^{(0)}(\lambda)$ and $B^{(1)}(\lambda)$ denotes the region to which they belong. 
Outside the sphere, $\psi_0(\mu,\nu)$ is 
an odd function with respect to $\nu$ 
so that we have $\psi_0(\mu,\nu)=-\psi_0(\mu,-\nu)$
and consequently $A^{(0)}(\lambda)=-B^{(0)}(\lambda)$ resulting in eq.\,(\ref{eqA:psi0}). 
Inside the sphere, 
$\psi_1(\mu,\nu)$ is finite everywhere including $\nu=+\infty$ 
($\nu\to+\infty$ denotes the smallest spherical surface, 
which approaches the Cartesian origin within the sphere $1$.)
Therefore, $A^{(1)}(\lambda)=0$ is necessary 
to keep the $e^{\lambda\nu}$ from blowing up the integrand.
Now, our object is to find the spectra $A^{(0)}(\lambda)$ and $B^{(1)}(\lambda)$ 
that match the boundary values of $\phi_0$ and $\phi_1$ 
on the interface $\nu=1$.

Two spectra $A^{(0)}(\lambda)$ and $B^{(1)}(\lambda)$ 
are determined by the two boundary conditions on $\nu=1$: 
(a) the continuity of potential across the boundary 
and (b) the discontinuity of normal component of electric field across the boundary,
which equals to the free surface charge density required by Gauss's law. 
Mathematically, we have 
\begin{subequations}
    \begin{align}
    \phi_0(\mu,1_-) &=\phi_1(\mu,1_+)
    \label{eqA:BCcontinuity}
    \\ 
    -(\mu^2+\nu^2)\left.\frac{\partial\phi_0}{\partial\nu}\right|_{\nu=1_-} 
    +&\er
    (\mu^2+\nu^2)\left.\frac{\partial\phi_1}{\partial\nu}\right|_{\nu=1_+}
    = \frac{1}{a}.
    \label{eqA:BCnormal}
    \end{align}
\end{subequations}
The normal component of electric field on the surface $\nu$ 
is $|E|=-(\mu^2+\nu^2)\frac{\partial\phi}{\partial\nu}$,
pointing inward (outward) on $\nu=1$ ($\nu=-1$).

To obtain the equations for the spectra, 
we manipulate eq.\,(\ref{eqA:BCcontinuity}) and eq.\,(\ref{eqA:BCnormal})
in following steps:
(1) Inserting eq.\,(\ref{eqA:potential});
(2) Dividing both sides by $(\mu^2+1)^{1/2}$; 
(3) Applying Hankel transformation of the zeroth order to both sides.  
Equation~(\ref{eqA:BCcontinuity}) is then transformed to
\begin{subequations}
    \begin{equation}
      2A^{(0)}(\lambda)\sinh\lambda
      =
      B^{(1)}(\lambda)e^{-\lambda}\equiv C(\lambda)
    \label{eqA:ClambdaA}
    \end{equation}
\end{subequations}
where $C(\lambda)$ is an auxiliary function defined for convenience. 
With eq.\,(\ref{eqA:ClambdaA}), 
two unknown spectra are effectively reduced to one. 
The normal boundary condition eq.\,(\ref{eqA:BCnormal}) 
is proceeded in the same way but dividing both sides by $(\mu^2+1)^{3/2}$.
In terms of $C(\lambda)$, it becomes 
\begin{equation}
    \begin{aligned}
    \int_0^\infty\!\dd\mu\,\frac{\mu J_0(\lambda\mu)}{\mu^2+1}
    \int_0^\infty\!\dd\tau\,\tau J_0(\tau\mu)C(\tau)
    -\frac{(\er+\coth\lambda)}{\er-1}\lambda C(\lambda)
    =
    \int_0^\infty\!\dd\mu\,
    \frac{\mu J_0(\lambda\mu)}{(\mu^2+1)(\mu^2+9)^{1/2}}
    -(e^{-3\lambda}+2e^{-\lambda}). 
    \end{aligned}
   \tag{8b} 
   \label{eqA:ClambdaB}
\end{equation}
Two useful Hankel transformations needed 
in our manipulation here are
\begin{subequations}
   \begin{align}
   \int_0^\infty \, 
   \mu\frac{\nu}{(\mu^2+\nu^2)^{1/2}}
   J_0(\lambda\mu)\dd\mu 
   &= e^{-\lambda\nu}/\lambda,     
   \label{eqA:hankel1}
   \\
   \int_0^\infty \, 
   \mu\frac{\nu}{(\mu^2+\nu^2)^{3/2}}
   J_0(\lambda\mu)\dd\mu 
   &= e^{-\lambda\nu}.     
   \label{eqA:hankel2}
   \end{align} 
\end{subequations}

The integral involving two Bessel functions $J_0$ 
on LHS of eq.\,(\ref{eqA:ClambdaB}) can be simplified 
by integrating over $\mu$ first.
The formula 6.541 in ref. 
\cite{gradshteyn2014table} 
provides us a clean result to it 
\begin{equation}
    \int_0^\infty\!\dd\mu\,\frac{\mu J_0(\lambda\mu)J_0(\tau\mu)}{\mu^2+1}
    = 
    \begin{cases}
        I_0(\lambda)K_0(\tau), \quad &\lambda\leq\tau,\\
        I_0(\tau)K_0(\lambda),\quad &\lambda>\tau. 
    \end{cases}
    \label{eqA:IK}
\end{equation}
where $I(x)$ and $K(x)$ are modified Bessel functions of the first kind and second kind, respectively. 
On the RHS of eq.\,(\ref{eqA:ClambdaB}), 
the integral can be simplied 
by inserting the inverse Hankel transform of $(\mu^2+9)^{-1/2}$
utlizing the eq.\,(\ref{eqA:hankel1}) and changig the order of integration 
\begin{equation}
    \int_0^\infty\!\dd\mu\,
    \frac{\mu J_0(\lambda\mu)}{(\mu^2+1)(\mu^2+9)^{1/2}}
    =
    \int_0^\infty\!\dd\tau\,
    e^{-3\tau}
    \int_0^\infty\!\dd\mu\,
    \frac{\mu J_0(\lambda\mu)J_0(\tau\mu)}{\mu^2+1}.
\end{equation}
The inner integral can be evaluated in terms of 
the modified bessel functions again 
by eq.\,(\ref{eqA:IK}). 

For future convenience, we define a new auxiliary function $g(\lambda)$ 
\begin{equation}
   g(\lambda) \equiv \frac{\er+\coth\lambda}{\er-1}\lambda C(\lambda).
   \label{eqA:glambda}
\end{equation}
Combining with the above simplifications on the integrals,
the integral equation of $C(\lambda)$ 
eq.\,(\ref{eqA:ClambdaB}) is eventually reduced to an integral equation of $g(\lambda)$
\begin{equation}
    \begin{aligned}
    g(\lambda) 
    -K_0(\lambda)\int_0^\lambda\!\dd\tau\, 
    I_0(\tau)\left(
    \frac{\er-1}{\er+\coth\tau}g(\tau) -e^{-3\tau}
    \right)
    -I_0(\lambda)\int_\lambda^\infty\!\dd\tau\, 
    K_0(\tau)\left(
    \frac{\er-1}{\er+\coth\tau}g(\tau) - e^{-3\tau}
    \right)
    =(e^{-3\lambda} + 2e^{-\lambda})
    \end{aligned}
    \label{eqA:glambdaIntEq}
\end{equation}

The integral equation of $g(\lambda)$ above is a Fredholm integral equation
of the second kind. 
It can be solved by transforming it to 
an differential equatino of $g(\lambda)$ 
after differentiating eq.\,(\ref{eqA:glambdaIntEq}) with respect to $\lambda$ twice. 
The first differentiation is carried out in the following steps:
(1) Dividing both sides by $K_0(\lambda)$;
(2) Differentiating both sides with respect to $\lambda$;
(3) multiplying both sides by $\lambda K_0^2(\lambda)$.
Equation\~(\ref{eqA:glambdaIntEq}) becomes 
\begin{equation}
    \begin{aligned}
    \lambda(K_0(\lambda)&g'(\lambda)+K_1(\lambda)g(\lambda)) 
    -\int_\lambda^\infty\!\dd\tau\, K_0(\tau)
    \left(\frac{\er-1}{\er+\coth\tau}g(\tau)-e^{-3\tau}\right) 
    =\lambda[-(3e^{-3\lambda}+2e^{-\lambda})K_0
    +(e^{-3\lambda}+2e^{-\lambda})K_1]
    \end{aligned}
    \tag{13.1}
    \label{eqA:glambdaIntEq1}
\end{equation}
The recurrence relation $K_0'(\lambda)=K_1(\lambda)$ 
and also the equivalence $K_0^2\left(\frac{I_0}{K_0}\right)'=1/\lambda$
have been used. 

Another direct differentiation with respect to $\lambda$ of 
eq.\,(\ref{eqA:glambdaIntEq1})
will get rid of the integral by Lebniz integral rule
and lead to an ODE of $g(\lambda)$ for asymmetric free charges ${\bf Q} = (q, -q)$
\begin{equation}
   \begin{aligned}
   (\lambda g')' 
   &+ \left(\frac{\er-1}{\er+\coth\lambda}-\lambda\right)g(\lambda)
   =(-2+8\lambda)e^{-3\lambda}, \\ 
   &g'(0) = -5, \quad g(\infty)=0,
   \end{aligned}
   \label{eqA:godeAsym}
\end{equation}
where another recurrence relation 
$K_1'(\lambda)=-K_0(\lambda)-K_1(\lambda)/\lambda$ has been used
and two boundary conditions are derived 
through examining the behavior of $g(\lambda)$ for $\lambda=0$ and $\lambda\to\infty$. 
Differentiating eq.\,(\ref{eqA:glambdaIntEq}) 
with respect to $\lambda$ and evaluating it at the limit $\lambda\to0$
will simply leave us with $g'(0)=-5$
as the derivatives of two integral expression 
with respect to $\lambda$ 
in eq.\,(\ref{eqA:glambdaIntEq}) vanish at $\lambda=0$. 
On the other hand, $g(\infty)=0$ is necessary 
to ensure the integrability of eq.\,(\ref{eqA:general}) when $\mu=0$.
Following the same procedure,
we can derive another ODE of $g(\lambda)$ and boundary conditions
for the case of symmetric free charge ${\bf Q} = (q, q)$. 
Here, we save the repetitive algebra and 
simply present its final result
\begin{equation}
    \begin{aligned}
       (\lambda g')' 
       &+ \left(\frac{\er-1}{\er+\tanh\lambda}-\lambda\right)g(\lambda)
       =-(-2+8\lambda)e^{-3\lambda}, \\ 
       &g'(0) = 5, \quad g(\infty)=0 .    
    \end{aligned}    
    \label{eqA:godeSym}
\end{equation}    
The major difference from eq.\,(\ref{eqA:godeAsym}) is that 
the hyperbolic cotangent function in the denominator 
becomes the hyperbolic tangent function, 
reflecting the even symmetry of potential about the $xy$-plane. 
The solutions of $g(\lambda)$ for both cases 
with several representative values of $\er$
are presented in Fig.\,(S1). 

The contact energy that we are looking for 
is simply $\mathcal{E}=\frac12 \sum_{i=1,2} Q_iV_i$ in our problem setting,
where $V_i$ is the mean surface potential of sphere $i$.
For spheres, the mean surface potential can be represented 
by the potential evaluated at its center due to the spherical symmetry.
Therefore, we only need the potentials at $(0,\pm 2,0)$ 
for the contact energies $\mathcal{E}=\frac12 Q_1\phi_1(0,2,0)+\frac12 Q_2\phi_2(0,-2,0)$,
in which $\phi_i$ is calculated by eq.\,(\ref{eqA:general}) 
with spectra found by solving respective ODE of $g(\lambda)$.
The contact energies for the dimer of identical spheres 
are presented for both cases of symmetric charges and asymmetric charges 
with $\er$ ranging from $0$ to $\infty$ in Fig.\,(S2).
The spectral method generalizes to 
dimer of spheres with different radii and permittivities immediately, 
and more importantly,
to spheres with higher order multipole charges.



\bibliographystyle{unsrt}
\bibliography{supplement}



\newpage 
\section*{Supplementary figures}

\renewcommand{\thefigure}{S\arabic{figure}}

\begin{figure}[h!]
    \centering
    \includegraphics[width=0.618\textwidth]{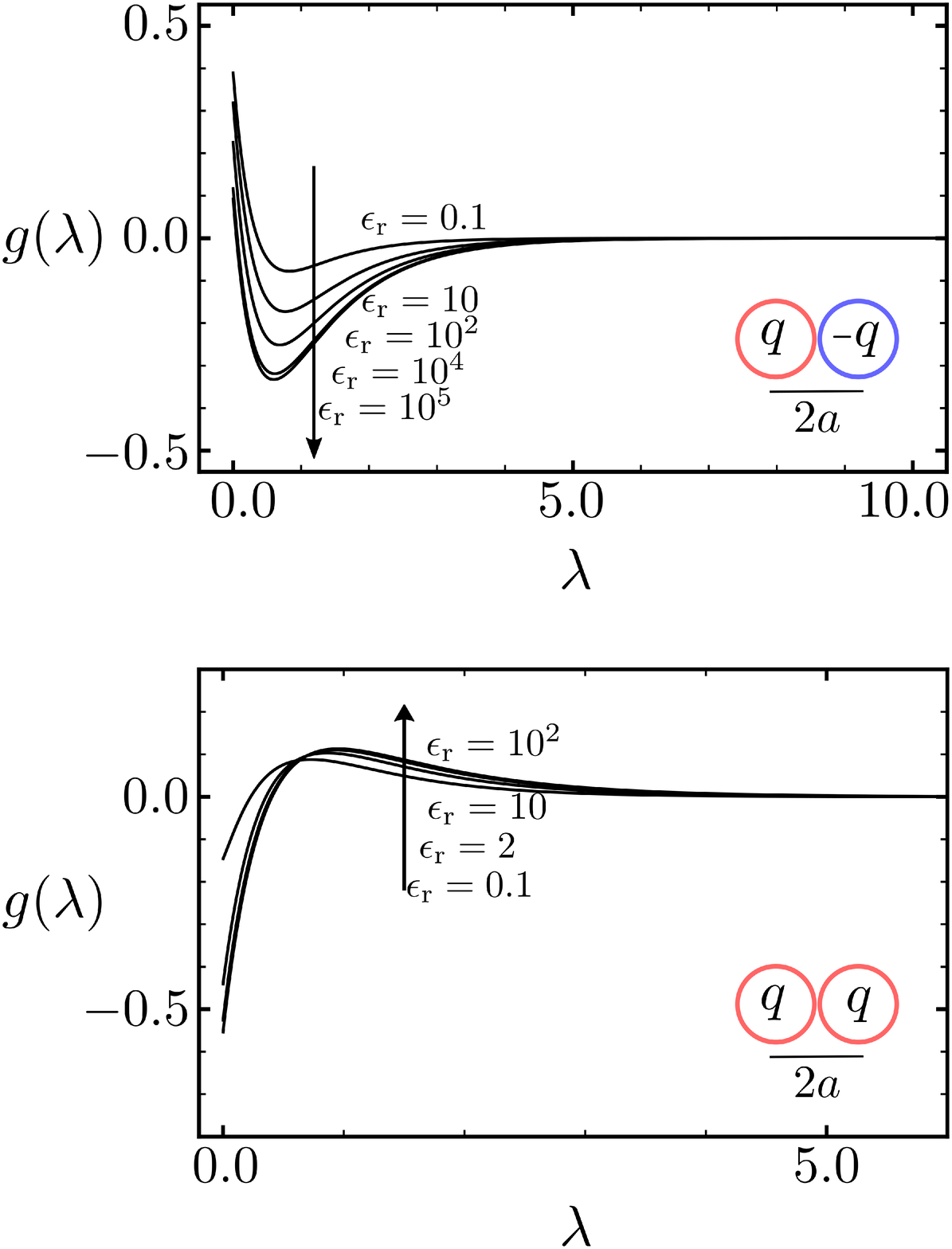}
    \caption{The solutions of $g(\lambda)$ for both cases 
    of asymmetric charges and symmetric charges 
    with several representative values of $\er$ are presented.
    The solutions converge slowly in the asymmetric case 
    due to the $\ln\er^2$ dependence of singular contact charges shown in the main text. 
    On the other hand,
    the ones converge fast in the symmetric case. 
    The solutions of $\er=10$ and $\er=10^2$ are 
    almost indistinguishable for the case of symmetric charges.
    }
    \label{fig:s1}
\end{figure}
\newpage 
\begin{figure}[h!]
    \centering
    \includegraphics[width=0.618\textwidth]{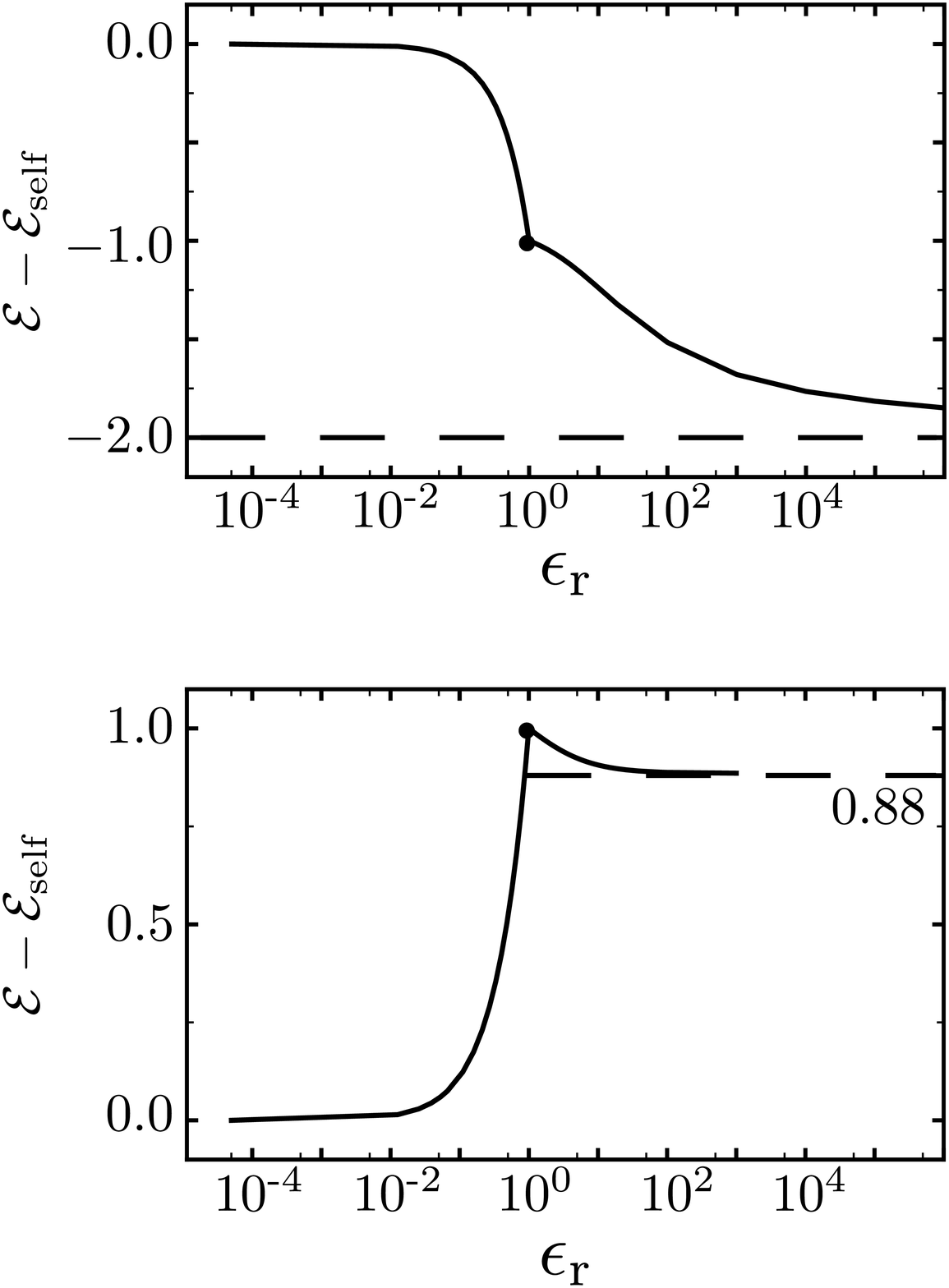}
    \caption{The electrostatic energy of a touching pair of identical spheres 
    for the case of asymmetric charges and symmetric charges.
    To emphasize the role of the electrostatic interaction,
    the energy of two isolated spheres, $\mathcal{E}_{\rm self}$
    is subtracted from the total energy $\mathcal{E}$.
    At $\er=1$, the polarization effect vanishes and 
    thus the interaction is simply the coulombic interaction 
    between two touching spehres, i.e. $-1$ and $1$,
    in the unit $q^2/(8\pi\epsilon_0a)$. 
    When $\er>1$, the polarization effect enhances the attraction
    and makes the interaction energy approach $-2$
    at the conducting limit $\er\to\infty$. 
    For the symmetric case, the interaction energy 
    quickly converges to the known result, $2/\ln2-2\approx0.88$.
    When $\er<1$, the polarization effect screens the coulombic interaction for both cases,
    which reduces the interaction energy to $0$ at about $\er\approx10^{-2}$. 
    }
    \label{fig:s2}
\end{figure}

\newpage 
\begin{figure}
    \centering
    \includegraphics[width=\textwidth]{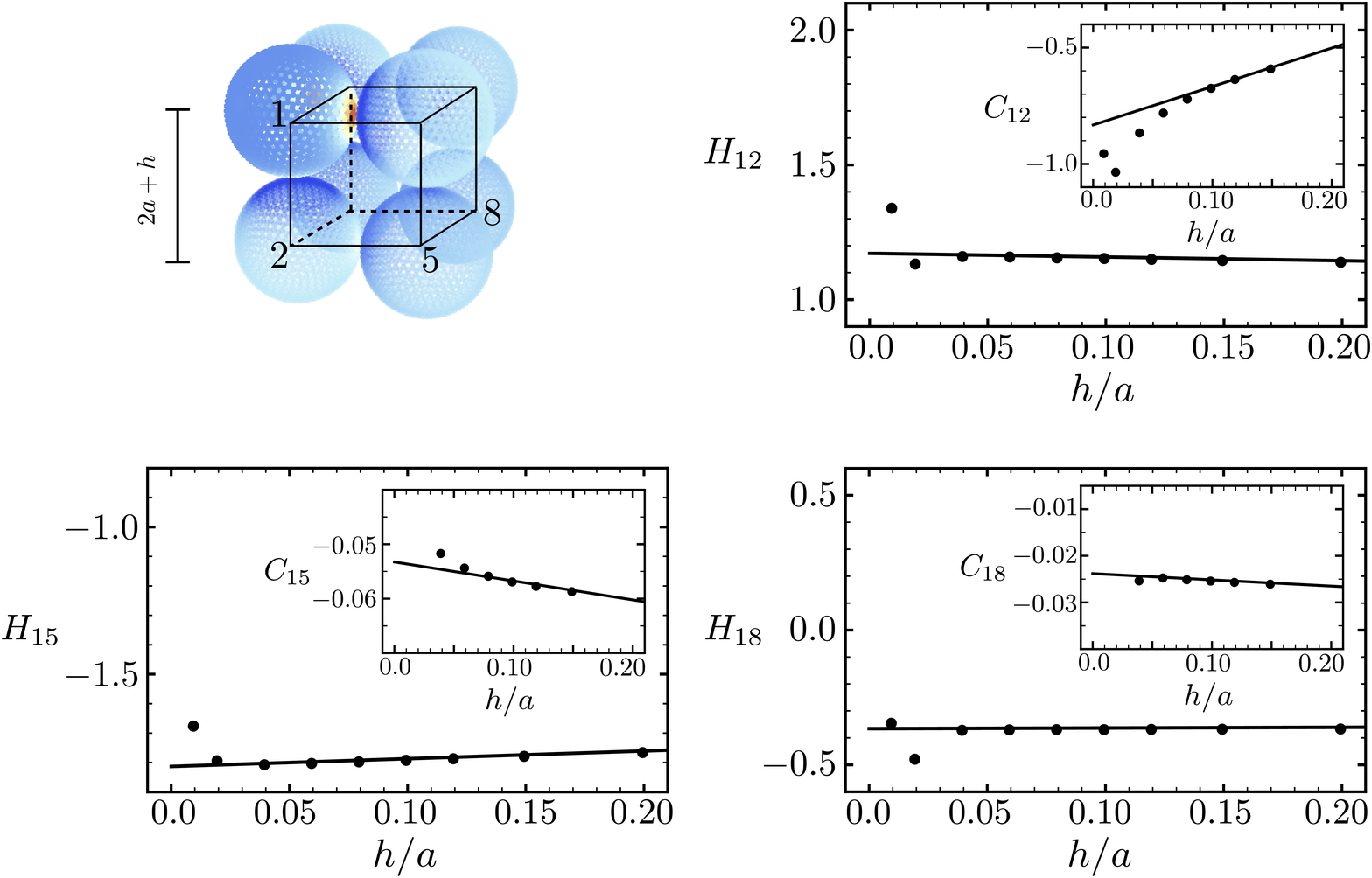}
    \caption{The regular part $H_{ij}$ of capacitance coefficient $C_{ij}$ 
    as a function of the normalized gap distance $h/a$.
    The indices of spheres are denoted in the sketch of the cube configuration.
    With respect to the sphere $1$, 
    sphere $2$ is one of the nearest neighbor, 
    sphere $5$ is on one of the plane-diagonal vertices,
    and sphere $8$ is on the body-diagonal vertex.
    The nonlinear behavior of $C_{ij}$
    gradually diminishes for the pair of spheres forming only a secondary contact. 
    }
    \label{fig:s3}
\end{figure}
\newpage 

\begin{figure}
    \centering
    \includegraphics[width=0.618\textwidth]{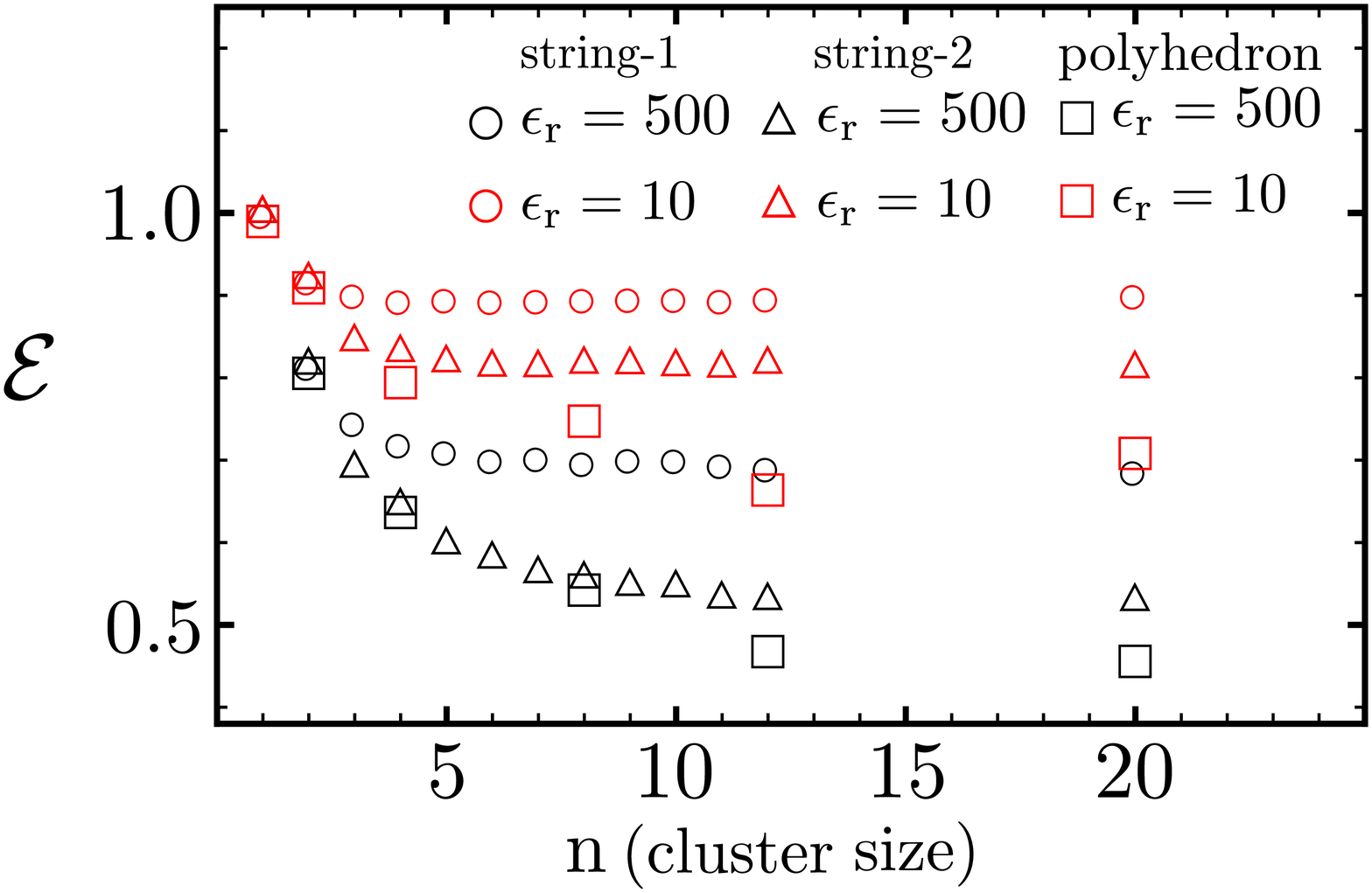}
    \caption{
    Size dependence of electrostatic energy $\mathcal{E}$ 
    for string-like and polyhedron packing
    with fixed total charge, when charge transfer is prohibited.
    The total charge $q$ resides on 
    one end of the string (string-$1$),
    the middle particle of the string (string-$2$),
    and an arbitrary vertex of polyhedron (polyhedron).
    }
    \label{fig:s4}
\end{figure}